# A novel real-time aeroelastic hybrid simulation system of section model wind tunnel testing based on adaptive extended Kalman filter

*Wenkai Du [a], Guangzhong Gao [a*], Suhan Li [a], Bo Fu [b], Jiawu Li [a], Ledong Zhu [c,d,e]*

[a] *School of Highway, Chang'an University, Xi'an, 710064, China.*
[b] *School of Civil Engineering, Chang'an University, Xi'an, 710064, China.*
[c] *Department of Bridge Engineering, Tongji University, 200092 Shanghai, China*
[d] *State Key Laboratory of Disaster Reduction in Civil Engineering, Tongji University, Shanghai, 200092, China*
[e] *Key Laboratory of Transport Industry of Bridge Wind Resistance Technology, Tongji University, Shanghai, 200092, China*

## Abstract

Elastically-supported section model tests are the most basic experimental technique in wind engineering, where helical springs are commonly employed to simulate the two-degree-of-freedom low-order modal motions of flexible structures. However, the traditional technique has intrinsic limitations in accurately modeling nonlinear structural behaviors and accurate adjustments of nonlinear structural damping. This study proposes a novel Real-Time Aeroelastic Hybrid Simulation system for section model wind tunnel tests by integrating an active control algorithm of adaptive Kalman filter. The proposed system enables the simulation of nonlinear heave-transverse-torsion coupled vibrations of a section model under the action of the oncoming wind. The structural properties, i.e. mass, damping and stiffness, are numerically simulated via an active control system, and the aerodynamic forces are physically modelled via the model-wind interaction in the wind tunnel. To validate the feasibility and accuracy of the proposed RTAHS system, a MATLAB/Simulink–FLUENT/UDF co-simulation framework is developed. Numerical verification results indicate that the proposed algorithm effectively estimates the motion responses in both linear and nonlinear scenarios.

## 1. Introduction

Section model wind tunnel testing is a fundamental method for studying the aerodynamic behaviors of slender flexible components in civil engineering, such as the main girder, hangers, stay cables, and transmission lines, etc. In the free vibration tests of a section model, the structural stiffnesses are commonly modelled via pre-tensioned helical springs, and such free vibration apparatus is conventionally employed to measure the wind-induced responses and evaluate the aeroelastic stability under the influence of unsteady aerodynamic forces. (Xu et al., 2023). Conventional free vibration setup of section model tests can only simulate the linear two-degree-of-freedom (DOF) oscillation of flexible structures (Gao et al., 2020; Gao et al, 2025). All structural properties, including mass, stiffness and damping are physically modelled by added mass blocks on the suspending arms, spring stiffness and external dampers, respectively, which makes it difficult and inefficient to accurately adjust the structural parameters. As a result, the structural properties of the model are often adjusted through a trial-and-error approach informed by researchers' expertise during wind tunnel tests (Chakraborty et al., 2022; Li et al., 2025; Li et al., 2024). Another limitation of conventional spring-suspended setup should also be highlighted, which is their inability to accurately model nonlinear structural behaviors, including both nonlinear stiffness and structural damping, whereas the modeling of nonlinear structural behaviors is essential for testing nonlinear wind-induced vibration of flexible structures. As a result, this limitation leads to inaccurate extrapolation of conclusions from scaled section models to full-scale structures (Gao and Zhu, 2015).

The method of real-time hybrid simulation (RTHS) has the potential to address all the above-mentioned limitations of conventional spring-suspended section model tests. The idea of real-time hybrid simulation was first introduced by Nakashima et al. (1992) and applied to structural seismic testing, and it has widespread application in civil engineering (Phillips and Spencer, 2013; Shao et al., 2016; Botelho et al., 2022; Imanpour et al., 2022). The general framework of RTHS is shown in Fig.1, and it involves treating the portion of the structure with complex and highly nonlinear behavior as the physical substructure for large-scale or full-scale testing (Pan et al., 2015). Meanwhile, the remaining parts of the structure, exhibiting weaker nonlinear behavior, are modelled as the numerical substructure for online simulation. By solving the differential motion equation of the numerical substructure, the displacement command at the boundary is determined, and real-time loading is applied to ensure accurate tracking of the displacement, velocity, and even acceleration responses. The reaction force from the physical substructure is fed back to the numerical substructure in real time, updating the displacement command, and the process continues in a feedback loop. Early RTHS had a maximum response frequency between 2 and 3 Hz and were applied to systems with more than 10 degrees of freedom (Nakashima and Masaoka, 1999).

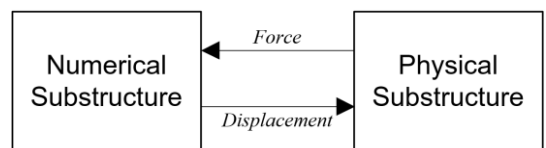

Fig. 1 General framework of real-time hybrid simulation

Real-time hybrid simulation has made significant advancements in seismic and wind engineering, driven by improvements in control algorithms and computational power. Yuan et al. (2017) conducted a real-time hybrid simulation of a typical multi-span continuous beam bridge and investigated the effects of various isolation bearings on the bridge's response under seismic motion. They demonstrated that the RTHS can accurately simulate the seismic response of seismic isolation bridges. Zhang et al. (2016) proposed a shake table RTHS framework for shear buildings and validated its effectiveness and robustness through numerical simulations. Kwon et al. (2019) and Moni et al. (2020) developed two hybrid simulation experimental setups and tested them with section models of a bridge deck and a high-rise building, respectively. Delay compensation and numerical integration schemes were employed to enhance experimental efficiency and stability. Wu and Song (2019) introduced a real-time aerodynamics hybrid simulation, using the reduced-scale aeroelastic mode of a tall building as the physical substructure and the full-scale dampers as the numerical substructure to accurately assess the wind-induced response of tall buildings. Wu et al. (2019) designed a novel controller that is entirely decoupled from the physical subsystem, enabling the effective simulation of both linear and nonlinear dynamic behaviors in flexible bridges. Lu et al. (2022) employed continuous- and discrete-time root locus techniques to predict and enhance the stability of multi-DOF real-time hybrid simulations. Dong et al. (2024) proposed an aeroelastic real-time hybrid simulation to study the vortex-induced vibration (VIV) of tall building structures.

Over the past two decades, real-time hybrid testing has matured in the simulations of low-DOF and uncoupled multi-DOF scenarios. However, limitations remain in the application of RTHS to section model wind tunnel tests. These limitations are summarized as follows: 1) The coupling accuracy between the physical and numerical models in RTHS still requires improvement, particularly for complex vibration responses involving relatively high-frequency dynamics and structural nonlinear effects. 2) Existing studies cannot be directly applied to the heave-transverse-torsion coupled vibrations of flexible structures under wind actions (Palacio-Betancur and Gutierrez Soto, 2023).

To resolve the above limitations, robust and high-precision control algorithms are needed when RTHS is applied to section model tests. In this connection, the algorithm of the Kalman filter (KF) is promising as it is well-suited for estimating state changes in various systems over time and has been widely applied in real-time signal processing, target tracking, and motion control (Khodarahmi and Maihami, 2023). Kalman filtering offers an optimal estimate of the system state by integrating the system model (prediction) with measurement data (update) (Kalman, 1960; Zhang et al., 2024). It is further extended to the Extended Kalman filter (EKF) for more complex nonlinear systems. At each step of system state estimation, the nonlinear system dynamics are linearized using the Taylor series expansion (Gustafsson and Hendeby, 2011). In the field of motion control, Palmieri et al. (2021) applied KF in collaborative robotics to accurately track human arm motion. Saito et al. (2020) presented an EKF for pose estimation, utilizing noise covariance matrices based on sensor output. Further research supports the effectiveness of the Kalman filter algorithm for motion tracking and control (Hu et al., 2015; Qian et al., 2023; Jin et al., 2023).

In this study, we aim to develop a novel real-time aeroelastic hybrid simulation system (RTAHS) for section model wind tunnel testing using an active control algorithm of adaptive extended Kalman filter (AEKF). The conventional spring-suspended system is replaced by a numerical substructure constructed via MATLAB/Simulink, where the AEKF is employed to estimate the model's motion states in real time, transferring the target displacement to the section model (physical substructure) to guide its vibration. Simultaneously, the physical model provides feedback by transmitting the measured aerodynamic force to the numerical substructure. To verify the feasibility and accuracy of the proposed method, computational fluid dynamics (CFD) is employed as the physical substructure and co-simulated with MATLAB/Simulink to validate both single-degree-of-freedom and two-degree-of-freedom systems. Additionally, the issue of time-delay associated with hybrid testing is discussed.

This paper is organized as follows: Section 2 derives the general governing equation of the RTAHS. Section 3 introduces the proposed RTAHS framework and experimental setup. Section 4 numerically validates the AEKF-RTAHS method in single-DOF and two-DOF vibration systems and presents a MATLAB/Simulink–FLUENT/UDF co-simulation framework. Section 5 discusses the time-delay issue. Finally, Section 6 presents the concluding remarks of this study.

## 2. Governing equation of RTAHS

Establishing the relationship between measured force and model motion is a key aspect of the RTAHS. For brevity, the governing equations are derived for the case of a heaving DOF vibration and the governing equations of a heave-transverse-torsion 3DOF system can be derived accordingly.

Firstly, the experimental setup of conventional spring-suspended section model tests is illustrated in Fig.2. The dynamic displacements are measured using displacement sensors targeted at the suspending arms and dynamic forces are measured using force sensors (not shown in Fig.2) installed at the ends of the section model and connected to the suspending arms through rid rods. To obtain the governing equations of RTAHS, the section model and the spring-suspension system are analyzed separately, as shown in Fig.2(b) and (c).

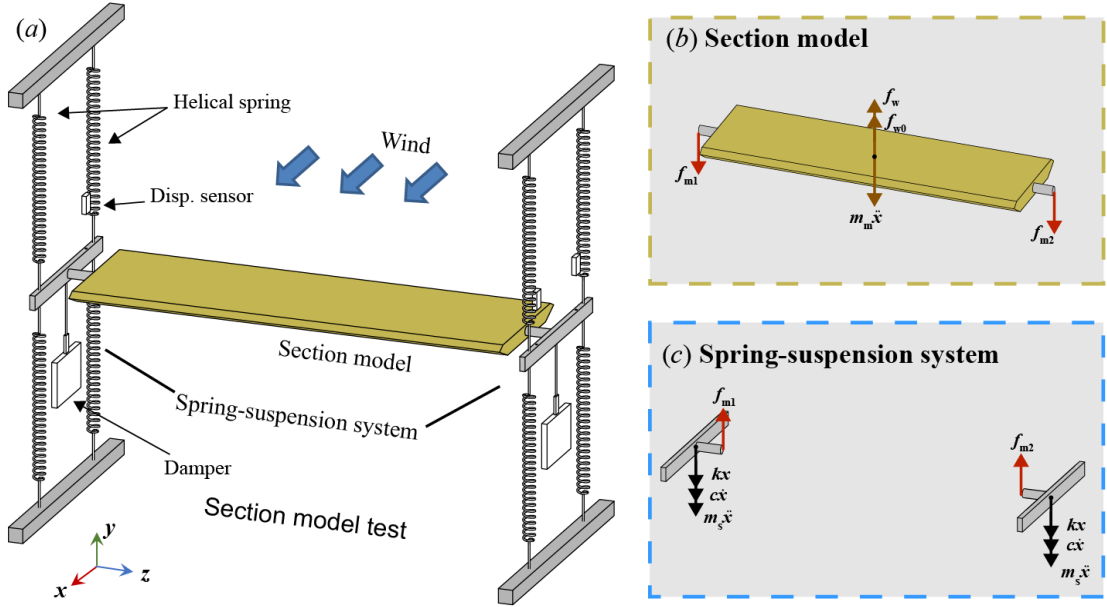

Fig. 2 Diagram of structural mechanical analysis in section model test: (a): schematics of section model test; (b): mechanical analysis of section model; (c): mechanical analysis of spring-suspension system

For the section model in Fig.2(b), its equilibrium equation can be rewritten as

$$f_w + f_{w0} - f_m - m_m \ddot{x} = 0 \tag{1}$$

where $f_{w0} = -m_a \ddot{x} - c_a \dot{x}$; $f_{w0}$ refers to the non-wind-induced added inertia and damping forces due to the interaction of section model and surrounding air in still air (Gao and Zhu, 2016), wherein $m_a$ and $c_a$ are the added mass and damping coefficient, respectively. $\ddot{x}$ is the acceleration response; $f_w$ is the wind-induced force in flowing air conditions; $f_m$ is the total measured force between the section model and spring-suspension system, $f_m = f_{m1} + f_{m2}$; $m_m$ represents the mass of the section model.

For the spring-suspension system in Fig.2(c), the equilibrium equation of suspending arms is

$$m_s \ddot{x} + c_s \dot{x} + kx - f_m = 0 \tag{2}$$

where $x$ and $\dot{x}$ are the displacement and velocity of the sectional model, respectively; $m_s$ is the equivalent mass of the spring-suspension system, which equals the total effective mass of the oscillatory system excluding the mass of the section model $m_m$. $c_s$ and $k$ are structural damping and spring stiffness, respectively.

The numerical substructure of RTAHS is used to replace the spring-suspension system in Fig.2(c) and only the section model is retained as the physical substructure to obtain accurate aerodynamic forces. Thus, Eq.(2) is the governing equation of RTAHS for the heaving-DOF oscillatory system. The targeted structural parameter (mass, damping ratio, frequency) could be set and adjusted in the numerical substructure, while the measured force is obtained in the wind tunnel.

For a heave-transverse-torsion coupled system, the governing equation can be expressed as

$$\mathbf{M}_s\ddot{\mathbf{X}} + \mathbf{C}_s\dot{\mathbf{X}} + \mathbf{K}\mathbf{X} - \mathbf{F}_m = \mathbf{0} \qquad (3)$$

where $\mathbf{M}_s = \mathrm{diag}\begin{bmatrix} m_{s,h} & m_{s,t} & m_{s,\alpha} \end{bmatrix}$, $\mathbf{C}_s = \mathrm{diag}\begin{bmatrix} 2m_{s,h}\xi_h\omega_h & 2m_{s,t}\xi_t\omega_t & 2m_{s,\alpha}\xi_\alpha\omega_\alpha \end{bmatrix}$, $\mathbf{K} = \mathrm{diag}\begin{bmatrix} m_{s,h}\omega_h^2 & m_{s,t}\omega_t^2 & m_{s,\alpha}\omega_\alpha^2 \end{bmatrix}$; $\mathbf{M}_s$, $\mathbf{C}_s$ and $\mathbf{K}$ are the matrixes of mass, structural damping, and stiffness of the spring-suspension system, respectively; the subscript $h$, $d$, $\alpha$ refer to heave, transverse and torsion DOFs, respectively; $\xi$ is the targeted structural damping ratio; $\omega$ is the circular frequency of each DOF; $\ddot{\mathbf{X}}$, $\dot{\mathbf{X}}$ and $\mathbf{X}$ are the vectors of acceleration, velocity and displacement of the section model, respectively; $\mathbf{F}_m$ is the vector of the measured dynamic forces for the vertical shear force, lateral shear force and torsional moment, respectively.

## 3. Real-Time Aeroelastic Hybrid Simulation system

In the following section, we develop a Real-Time Aeroelastic Hybrid Simulation (RTAHS) for testing the complex wind-induced behaviors. The aerodynamic forces depend on the aerodynamic shape of tested model in the wind tunnel, which is difficult to describe numerically, and are therefore modelled by the physical substructure, i.e., the section model. The elastically-supported system, governed by Eq.(2), serves as the numerical substructure. This section introduces the framework of RTAHS and its adaptive extended Kalman filter algorithm (AEKF). Subsequently, the principle of AEKF in MATLAB/Simulink will be discussed. Finally, the physical substructure of RTAHS enabling heave-transverse-torsion 3DOF vibration will be presented in detail.

### 3.1. RTAHS framework

As illustrated in Fig.3, the RTAHS framework consists of two parts: the numerical substructure and the physical substructure. The numerical substructure is constructed in Matlab/Simulink, while the physical substructure is installed in the wind tunnel. Initially, the dynamic force signals are measured via force sensors for the section model and then transmitted to the numerical substructure. Using the pre-determined structural parameters, the dynamic control equations are numerically solved at each step to obtain the desired dynamic displacement. The next step involves Kalman filter processing, which serves as the core module of the RTAHS. After fusing the model's observation data (measured signals of dynamic force and displacement), the motion state of the model for the next time step is predicted, and the corresponding displacement command is transmitted to the motor control device in the physical substructure to guide the motor rotation. Time-delay compensation is considered at this stage. Finally, the sensors feedback the measurement data, recording the vibration response and initiating the next cycle. Throughout this process, each component is closely interconnected to monitor and control the motion state of the model in real time.

Compared with the traditional technique of spring-suspended section model tests, the proposed approach of RTAHS offers several advantages. Firstly, the structural parameters such as mass, structural damping and stiffness can be independently input and adjusted in the numerical substructure, and thus only the accurate modelling of the aerodynamic shape is required, thereby significantly simplifying the experimental process. More importantly, the nonlinear structural behaviors of flexible bridges can be conveniently incorporated by simply adjusting the structural parameters in the numerical substructure, resulting in more realistic and accurate experimental results. Furthermore, the approach can be extended to investigate 3-DOF heave-transverse-torsion coupled vibrations for flexible components such as suspension cables, transmission lines, etc.

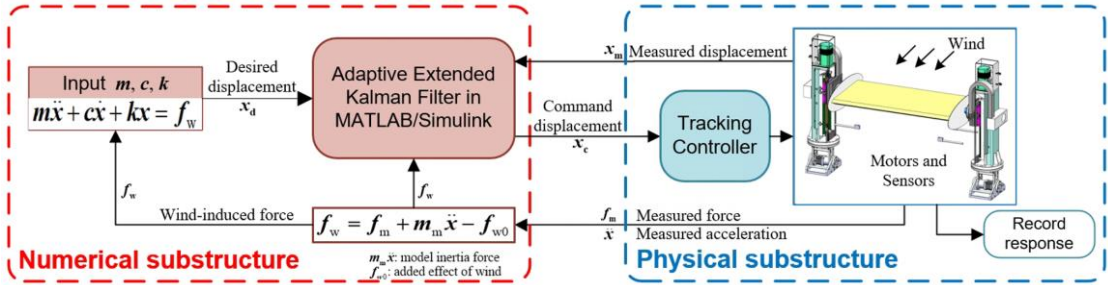

Fig. 3 Framework of the proposed RTAHS

### 3.2. Numerical substructure: control algorithm

#### 3.2.1 Adaptive Extended Kalman Filter

Kalman filter is a recursive digital processing algorithm widely used for real-time state estimation of linear dynamic systems (Urrea and Agramonte, 2021; Welch, 2021). It recursively updates the state estimate using prior knowledge and measurement data, providing the best prediction of the future state. In cases where the system exhibits nonlinear dynamics or nonlinear observation models, or where non-Gaussian noise is present, the extended Kalman filter (EKF) is employed.

However, both the KF and EKF assume that the process noise and measurement noise are white noise with zero mean and known statistical characteristics. As a result, the filtering algorithm is prone to error accumulation, which can lead to the error covariance matrix losing positive definiteness or symmetry. Additionally, the EKF is sensitive to numerical instability. To address these limitations, an improved adaptive extended Kalman filter (AEKF), also known as the maximum posterior estimator, is employed in the present study. This approach aims to deduce the statistical characteristics of process noise and measurement noise from the updated measurements and optimize the filter performance. The detailed derivation of the AEKF formula is provided in the Appendix.

The AEKF is employed to estimate the model state and then guides the tracking controller. In the proposed RTAHS, the governing equation Eq.(3) is incorporated into the state transition equation of the AEKF in order to estimate the prior state, which is expressed as Eq.(4a). Eq.(4b) is the observation equation incorporating measurement noise, which is expressed as

$$\dot{x} = Ax + Bu + w \tag{4a}$$

$$y = H_m x + v \tag{4b}$$

where $\mathbf{x}=\begin{bmatrix} x_v & x_h & x_\alpha \end{bmatrix}^T$ and $\dot{\mathbf{x}}$ are the augmented state vectors of the numerical model, being $\mathbf{x}_i=\begin{bmatrix} x_i & \dot{x}_i \end{bmatrix}(i=v,h,\alpha)$; the subscript $v, h, \alpha$ refer to vertical, horizontal and torsional, respectively; $\mathbf{u}$ is the input vector, which is relative to wind-induced force; $\mathbf{y}$ is observation vector of the physical model; $\mathbf{H}_m = \mathrm{diag}\begin{bmatrix} \mathbf{H} & \mathbf{H} & \mathbf{H} \end{bmatrix}$ is observation matrix with $\mathbf{H}=\begin{bmatrix} 1 & 0 \end{bmatrix}$; $\mathbf{w}$ represents the system process noise vector with covariance matrix $\mathbf{Q}$; $\mathbf{v}$ represents the measurement noise vector with the covariance matrix $\mathbf{R}$. The matrixes $\mathbf{A}$ and $\mathbf{B}$ are the state matrix and input matrix, respectively, which are expressed as

$$\mathbf{A}=\begin{bmatrix} \Lambda_v & & \\ & \Lambda_h & \\ & & \Lambda_\alpha \end{bmatrix}; \mathbf{B}=\begin{bmatrix} \Gamma_v & & \\ & \Gamma_h & \\ & & \Gamma_\alpha \end{bmatrix}$$

where

$$\Lambda_i = \begin{bmatrix} 0 & 1 \\ \omega_i^2 & -2\xi_i\omega_i \end{bmatrix}(i=v,h,\alpha); \; \Gamma_v = \begin{bmatrix} 0 \\ 1/m_v \end{bmatrix}; \; \Gamma_h = \begin{bmatrix} 0 \\ 1/m_h \end{bmatrix}; \; \Gamma_\alpha = \begin{bmatrix} 0 \\ 1/I \end{bmatrix}$$

*3.2.2 AEKF in MATLAB Simulink*

In Simulink, the KF and EKF modules have been encapsulated, allowing for easy and rapid adoption. Whereas, the AEKF must be developed through a function module with broader applicability. Fig. 4 illustrates the designed framework of the AEKF module in Simulink, where the corresponding equations are derived in the Appendix. The AEKF module contains four steps: measurement, update, prediction, and adaptive adjustment. First, the initial state and uncertainty of the section model oscillatory system are estimated and input, and the sign of the measured state and uncertainty are specified. Next, the Kalman gain is calculated according to Eq.(A-12). Then, the current state is estimated using Eq.(A-13), and the estimated uncertainty is updated by Eq.(A-14) accordingly. In the third step, the predicted state for the next iteration is calculated using the system's dynamic model and Eq.(A-10), and the estimated uncertainty is extrapolated by Eq.(A-11). In the final step, as shown in Eqs.(A-15)-(A-19), the state covariance and measurement covariance are adaptively adjusted based on the innovation, which is the difference between the measured value and predicted state. Thus, the accuracy of the update and prediction steps is influenced by the adaptive adjustment, which plays a crucial role in the AEKF.

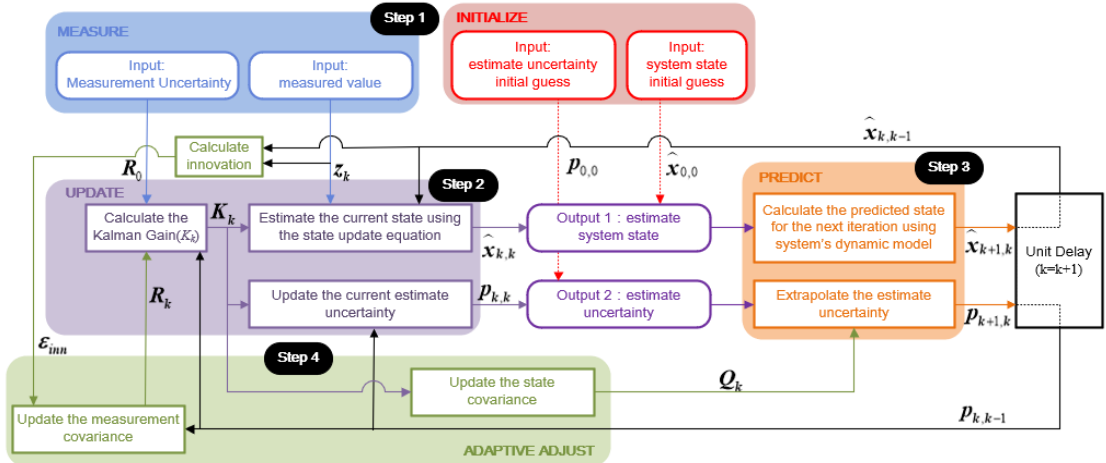

Fig. 4 The procedure of the AEKF module in MATLAB/Simulink

### 3.3. Physical substructure: experimental setup

The physical substructure consists of four servo motors, two joint motors, two reducers, sliding rails mounted to the upright, ball screws, drag chains, and various sensors, as shown in Fig. 5. These components are securely fixed to the structural upright and enclosed within a transparent plastic casing. The overall height of the apparatus is 1.2 m, with a movable range of 0.4 m for the sliding segment, which adequately meets the experimental requirements for flexible structural elements.

The oscillation of a section model is achieved through the coordinated operation of the servo motors and joint motors. In this study, all servo motors are driven by target displacement control. The six motors are synchronized and controlled by an external track controller module. They are driven by servo drives with high pulse counts and high-resolution encoders, allowing for accurate control of micro-movements in vibration. The servo system is equipped with a real-time feedback mechanism that continuously monitors the position and velocity of the output shaft, ensuring high-precision motion control. The joint motors could be equipped with 17-bit encoders, achieving a torsional motion resolution of 0.003°. Compared to stepper motors, servo motors provide faster motion response, superior overload capacity, and lower noise.

The servo motor generates rotational motion through electromagnetic induction, while the reducer employs multi-stage gear reduction to provide the required low-speed, high-torque output shaft. This approach reduces the inertial impact of the load on the motor and enhances the system's control precision. Two parallel guide rails are fixed to the upright, offering auxiliary support to the model. Frictional losses generated during the vibration process are compensated by increasing the driving force of the servo motor. The ball screw converts rotational motion into linear motion utilizing rolling balls, which require relatively low starting torque to enable precise micro-feeding. Compared to sliding screw mechanisms, ball screws offer higher precision and transmission efficiency. Similar to the drag chain in machine tools, the drag chain moves vertically with the model's motion, protecting the wiring of the joint motors that pass through it. In wind tunnel tests, the force signals are measured using bar-type force balances, which are installed at the two ends of the section model and connected to rigid arms. Laser displacement sensors and piezoelectric accelerometers are used to measure the model's displacement and acceleration, respectively.

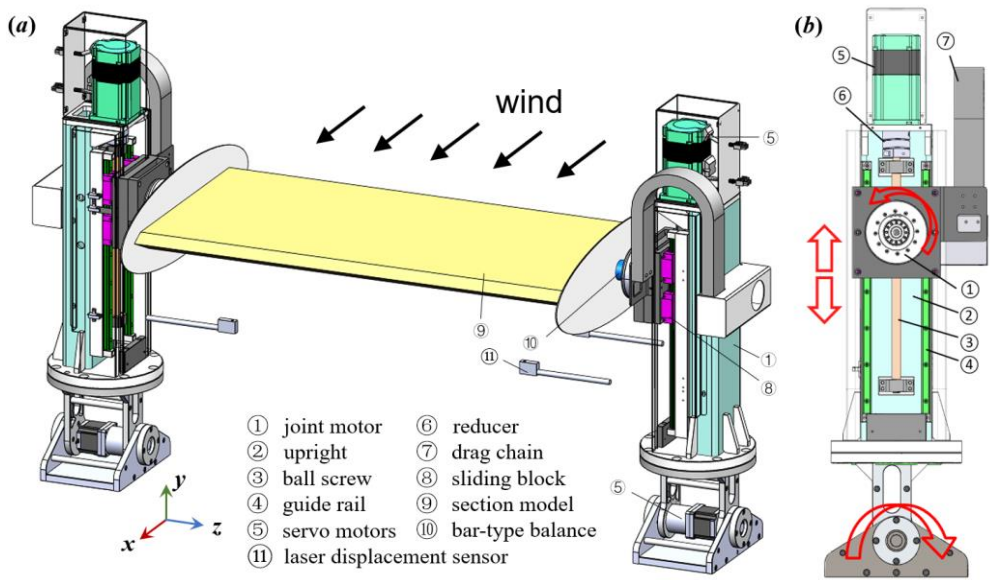

Fig. 5 The experimental setup of the proposed RTAHS: (a) schematic diagram of mechanical apparatus; (b) front view of the experimental setup.

For the vertical motion, the servo motor at the top achieves the vertical movement of the model by rotating the ball screw. The joint motors, which are in relatively smaller sizes and located at both ends of the model, are responsible for the torsional angle of the section model. For the transverse motion, motion compensation should be calculated and applied to the motor's target displacement. As is illustrated in Fig. 6, when the model moves from position A to position B, lateral displacement $\delta H$ and vertical displacement $\delta V$ occur simultaneously:

$$\delta V = V(t+\delta t) - V(t) \tag{5a}$$

$$\delta H = H(t+\delta t) - H(t) \tag{5b}$$

According to geometric relations, the target displacements of the top and bottom servo motors are:

$$\delta l = [H(t+\delta t)^2 + V(t+\delta t)^2]^{1/2} - [H(t)^2 + V(t)^2]^{1/2} \tag{6a}$$

$$\delta \alpha = \arctan[H(t+\delta t)/V(t+\delta t)] - \arctan[H(t)/V(t)] \tag{6b}$$

Meanwhile, the section model will experience additional rotation due to the motion of the bottom motor. To compensate for this, the joint motor should rotate in the opposite direction by the same displacement $\delta \alpha$. Therefore, for a 3DOF heave-transverse-torsion coupled system, the target displacement of the joint motor is:

$$\delta \beta = \delta \alpha_0 - \delta \alpha \tag{7}$$

where $\delta \alpha_0$ is the target torsional displacement of the section model.

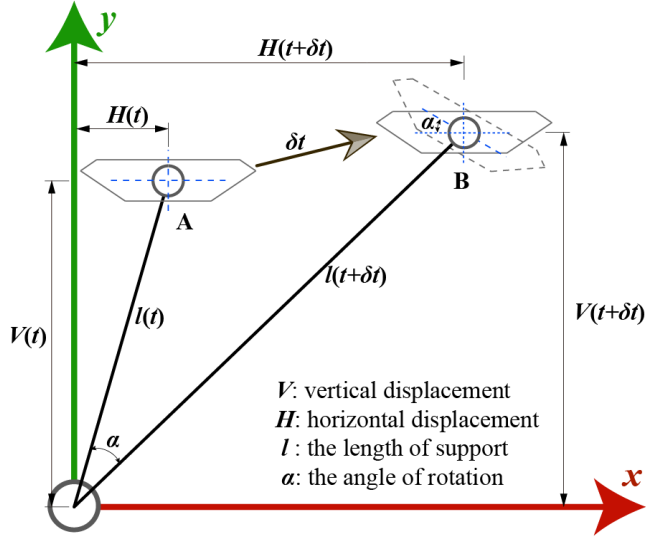

Fig. 6 Motion compensation of heave-transverse-torsion coupled system

## 4. Numerical validation

The wind-induced force $f_w$ of Eq.(1) can be directly calculated from CFD. The control equation for the validation part can be derived by subtracting Eq.(2) from Eq.(3):

$$m\ddot{x} + c\dot{x} + kx = f_w \tag{8}$$

where $m = m_m + m_s + m_a$ and $c = c_s + c_a$.

For a 3DOF heave-transverse-torsion coupled system, the equation can be expressed as

$$\mathbf{M}\ddot{\mathbf{X}} + \mathbf{C}\dot{\mathbf{X}} + \mathbf{K}\mathbf{X} = \mathbf{F}_{se} \tag{9}$$

where $\mathbf{M}$, $\mathbf{C}$ are the mass matrix and damping matrix, which consist of the section model system and spring-suspension system shown in Fig.2; $\mathbf{K}$ is the stiffness matrix of the spring-suspension system; $\mathbf{F}_{se}$ is the vector of the aeroelastic forces.

To evaluate the performance of the RTAHS, time history analysis was conducted for three vibration systems, i.e., a single-DOF linear system, a single-DOF nonlinear system, and a two-DOF heave-torsion coupled system. For each case, Eq.(9) is set as the governing equation for the numerical substructure, which aims to predict and guide the motion state of the model. The performance of the RTAHS was assessed by comparing the displacement signals from the RTAHS and other high-fidelity numerical methods. The details of the numerical validations are listed in Table 1. Notably, the co-simulation framework of MATLAB Simulink and ANSYS Fluent was developed in the numerical validation of the heave-torsion coupled system.

Table 1 The Setup of numerical validation cases

| Case | RTAHS | | Validation method |
|---|---|---|---|
| | Numerical substructure | Physical substructure | |
| single-DOF linear system | Kalman filter (KF) | Newmark-β method | Newmark-β method |
| single-DOF nonlinear system | Extend Kalman filter (EKF) | 4th order Runge-Kutta method | 4th order Runge-Kutta method |
| two-DOF heave-torsion coupled system | Adaptive Extend Kalman filter (AEKF) | ANSYS Fluent | CFD |

## 4.1. Case I: single-DOF linear system

The proposed RTAHS was first tested on single-DOF linear vibration. The heaving instability which is the most common in engineering, was taken as the validation scenario. The dynamic parameters are based on the wind tunnel tests by Zhu et al. (2013). The dynamic equivalent mass of the suspended mass lump system ($m$) was calculated as 182.178 kg, accounting for one-third of the mass of the springs and the mass of additional accessories such as screws. The linear self-excited force model is simplified as Eq.(11). The governing equation for vertical vibration is as follows:

$$m(\ddot{h} + 2\xi_{h0}\omega_{h0}\dot{h} + \omega_{h0}^2 h) = f_{VI}(t) \quad (10)$$

$$f_{VI}(t) = (1/2)\rho U^2 (2D)[Y_1(\dot{h}/U) + Y_2(h/U)] \quad (11)$$

where $\ddot{h}$, $\dot{h}$ and $h$ are the heaving acceleration, velocity and displacement of the section model, respectively; $\xi_{h0} = 0.005$ is the damping ratio; $\omega_{h0} = 17.64$ rad/s is the natural circular frequency; $f_{VI}(t)$ is the vortex-induced vertical force; $U = 9.1$ m/s is the wind speed; $D = 0.175$ m refers to the height of the section model; $\rho = 1.25$ kg/m³ is the air density; $Y_1$ and $Y_2$ are the reduced frequency-depended coefficients of aeroelastic self-excited force; all the parameters above is obtained from Zhu et al. (2013).

The signal of self-excited force is calculated using the Newmark-beta method in MATLAB and then delivered to the numerical substructure of RTAHS. The *Kalman filter* block provided by Simulink was adopted in the linear system. The matrices $\mathbf{A}$, $\mathbf{B}$ and $\mathbf{H}_m$ in the KF block are defined below. In this case, the process noise and measurement noise are random with the same noise covariance $10^{-5}$.

$$\mathbf{A} = \begin{bmatrix} 0 & 1 \\ \omega_{h0}^2 & -2\xi_{h0}\omega_{h0} \end{bmatrix}; \mathbf{B} = \begin{bmatrix} 0 \\ 1/m \end{bmatrix}; \mathbf{H}_m = \begin{bmatrix} 1 & 0 \end{bmatrix}$$

Fig. 7 shows the calculation results of the RTAHS method and the Newmark-β method. It is evident that the displacement curves obtained using the RTAHS alignment are in good agreement with the Newmark-β method, regardless of whether the system is in convergence or divergence states. This indicates that the proposed RTAHS is effectively applicable to the single-DOF linear system of section model tests.

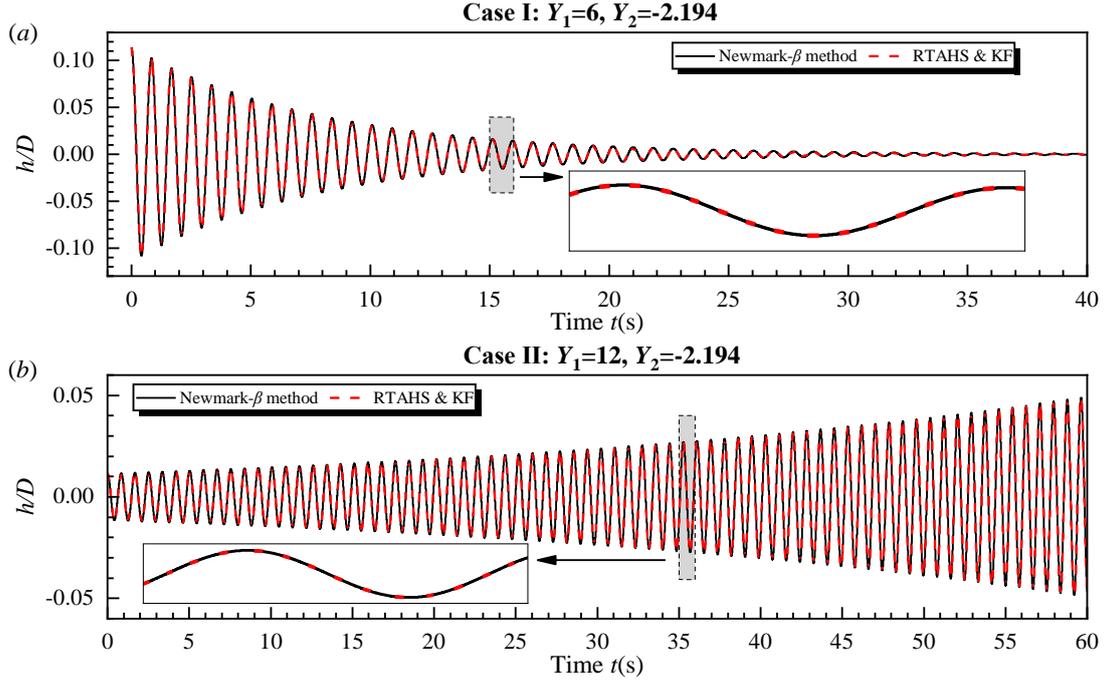

Fig.7 Numerical validation of a single-DOF linear system: (a) Case I: $Y_1 = 6.5$, $Y_2 = -2.194$; (b) Case II: $Y_1 = 11.966$, $Y_2 = -2.194$.

## 4.2. Case II: single-DOF nonlinear system

To evaluate the RTAHS in the single-DOF nonlinear system, the physical substructure utilizes the fourth-order Runge-Kutta method to iteratively calculate the model motion with measurement noise. The signals of displacement and dynamic forces are then input into the EKF block of the numerical substructure to estimate the motion state at each time step, and the results are compared with those obtained from the fourth-order Runge-Kutta method. Specifically, Scanlan's empirical nonlinear vortex-force model is used to calculate the nonlinear self-excited force signals (Zhu et al., 2013), while the nonlinear damping ratio and frequency functions, which depend on transient amplitude, are obtained from Gao et al. (2015). These equations are provided below.

The governing equation for the single-DOF nonlinear system is

$$m[\ddot{h} + 2\xi(a)\omega(a)\dot{h} + \omega(a)^2 h] = f_{VI}(h, \dot{h}) \qquad (12)$$

where $\xi(a)$ and $\omega(a)$ refer to the amplitude-dependent damping and circular frequency.

The nonlinear self-excited force model is expressed as

$$f_{\text{VI}}(h,\dot{h}) = \frac{1}{2}\rho U^2 (2D)[Y_1(1-\varepsilon\frac{h^2}{D^2})\frac{\dot{h}}{U} + Y_2\frac{h}{D} + \frac{1}{2}\tilde{C}_L \sin(\omega_{vs}t+\psi)] \quad (13)$$

where $Y_1$, $Y_2$ and $\varepsilon$ are the reduced frequency-depended coefficients of aeroelastic self-excited force; $\tilde{C}_L$ is the amplitude of the vortex-shedding force; $\omega_{vs}$ is the circular frequency of vortex shedding and usually in the vicinity of $\omega_{h0}$ during vortex-induced vibration; $\Psi$ represents the phase difference between the vortex shedding and the displacement response.

The nonlinear structural parameters in the numerical substructure are defined as (Gao and Zhu, 2015)

$$\xi(a) = 1.247\times 10^{-4}/(2a/D) + 3.65\times 10^{-3} + 1.264\times 10^{-2}\cdot(2a/D) \quad (14)$$

$$\omega(a) = \omega_0(1-a/5D) \quad (15)$$

where $a(t) = [h(t)^2 + (\dot{h}(t)/\omega_0)^2]^{1/2}$ is instantaneous heaving amplitude; $\omega_0$ is the natural circular frequency.

Compared to the Kalman filter algorithm, the EKF is more effective at capturing nonlinear vibration responses, as discussed in Section 3. In Simulink, the *Extended Kalman Filter* block estimates the states of a discrete-time nonlinear system using the first-order discrete-time extended Kalman filter algorithm. The nonlinear state transition function and measurement functions for the system should be created and specified within the block. Other parameters remain consistent with those of Case I. Additionally, the block supports state estimation for systems with multiple sensors operating at different sampling rates. The relevant dynamic parameters are listed in Table 2.

Table 2 Dynamic parameters of the 1DOF nonlinear system

| Aerodynamic Parameters | Values | Parameters of EKF | Values |
|---|---|---|---|
| $Y_1$ | 6.5 and 11.966 | Initial state covariance | $10^{-10}$ |
| $Y_2$ | -2.194 | Process noise $\hat{q}$ | 0 |
| $\varepsilon$ | 0.5 | Process covariance $Q$ | $10^{-8}$ |
| $\tilde{C}_L$ | -0.022 | Measurement noise $\hat{r}$ | 0 |
| $\omega_{vs}$ | 0.4477 | Measurement covariance $R$ | $10^{-8}$ |
| $\Psi$ | -0.0128 | - | - |

Fig. 8 illustrates the motion time history for two typical conditions of aerodynamic damping, i.e., positive damping and negative damping. It is observed that the calculated displacement from the proposed RTAHS aligns well with those obtained from the fourth-order Runge-Kutta method. In addition, the performance of KF and EKF for nonlinear systems are also compared, as shown in Fig. 9., it is clear that the application of the EKF method is necessary to guarantee the accuracy of simulated nonlinear responses. As a result, the proposed RTAHS is validated to be well-suitable for the single-DOF nonlinear vibration system of section model tests.

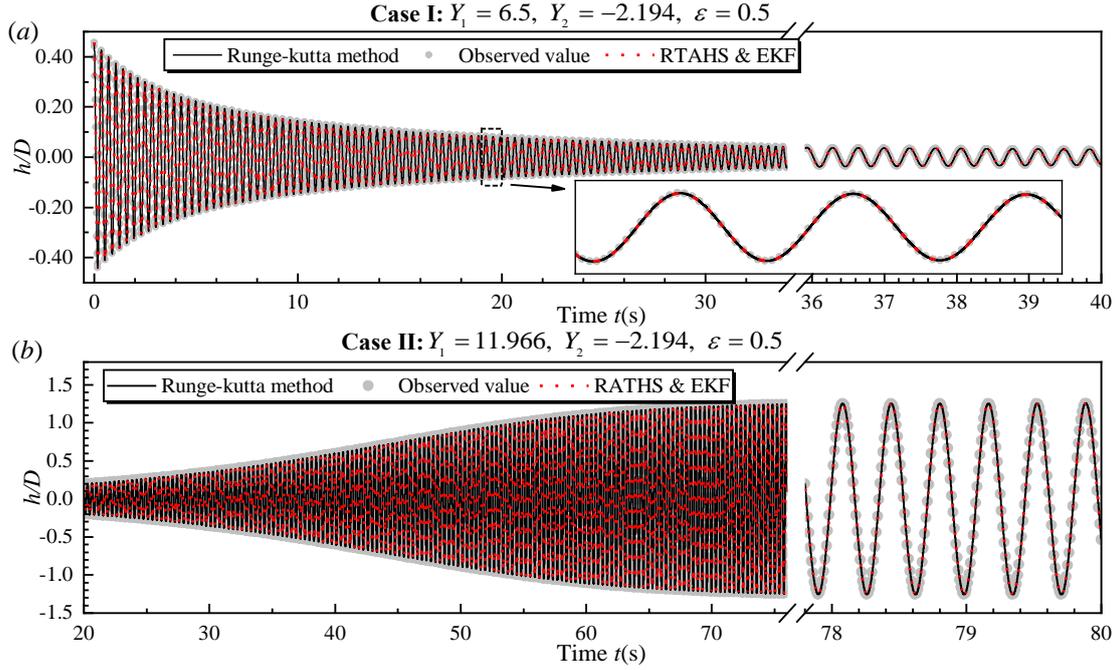

Fig. 8 Numerical validation of a single-DOF nonlinear system: (a) Case I: $Y_1 = 6.5$, $Y_2 = -2.194$, $\varepsilon = 0.5$; (b) Case II: $Y_1 = 11.966$, $Y_2 = -2.194$, $\varepsilon = 0.5$.

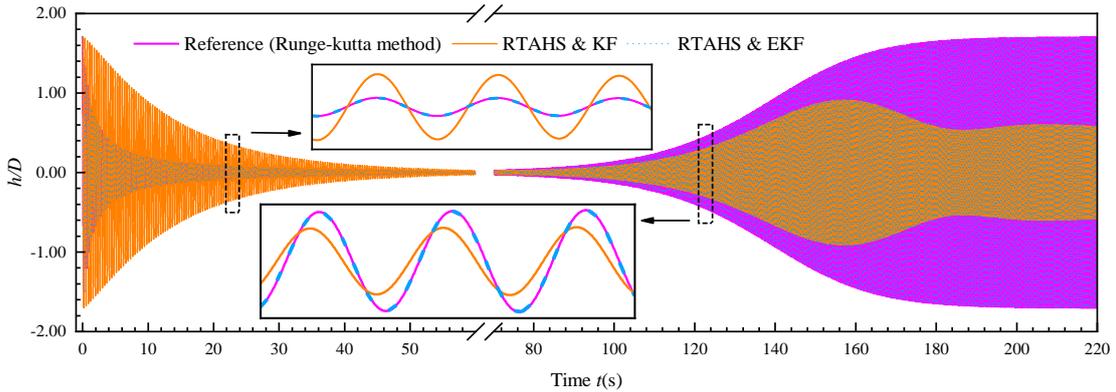

Fig. 9 Comparison of EKF-RTAHS and KF-RTAHS in a single-DOF nonlinear system.

### 4.3. Case III: two-DOF heave-torsion coupled system

For the two-DOF heave-torsion coupled system, Computational Fluid Dynamics (CFD) was employed as the physical substructure within a co-simulation framework with MATLAB/Simulink. In this way, a full feedback loop is formed between the numerical substructure and physical substructure, which is the case of free vibration during structure-wind interaction. The dynamic force was calculated via CFD in each time step, which is equivalent to the measured dynamic force in the section model test and then transferred to the numerical substructure (MATLAB/Simulink) to predict the vibration response of the next step.

*4.3.1 MATLAB/Simulink – FLUENT/UDF co-simulation framework*

To validate the proposed RTAHS more rigorously, a MATLAB/Simulink – FLUENT/UDF co-simulation framework was developed to simulate a full feedback loop. The flowchart of the co-simulation framework is illustrated in Fig.10. In this framework, MATLAB/Simulink and ANSYS Fluent perform simultaneously and exchange data in real time. On the Simulink side, the structural parameters **M**, **C**, and **K** are defined initially. Simulink then receives the signals of aerodynamic force and displacement from ANSYS Fluent at each time step, and then performs numerical calculation via AEKF, and sends the displacement command signal to the Fluent side. Thus, Simulink serves as the numerical substructure in the RTAHS. On the Fluent side, the force and displacement signals of the model are monitored in real-time and transmitted to Simulink. Simultaneously, the model's motion is guided, and the dynamic mesh area is updated based on the feedback target response. Therefore, the ANSYS Fluent acts as the role of the physical substructure in the RTAHS, just like the section model to receive the aerodynamic force during structure-wind interaction.

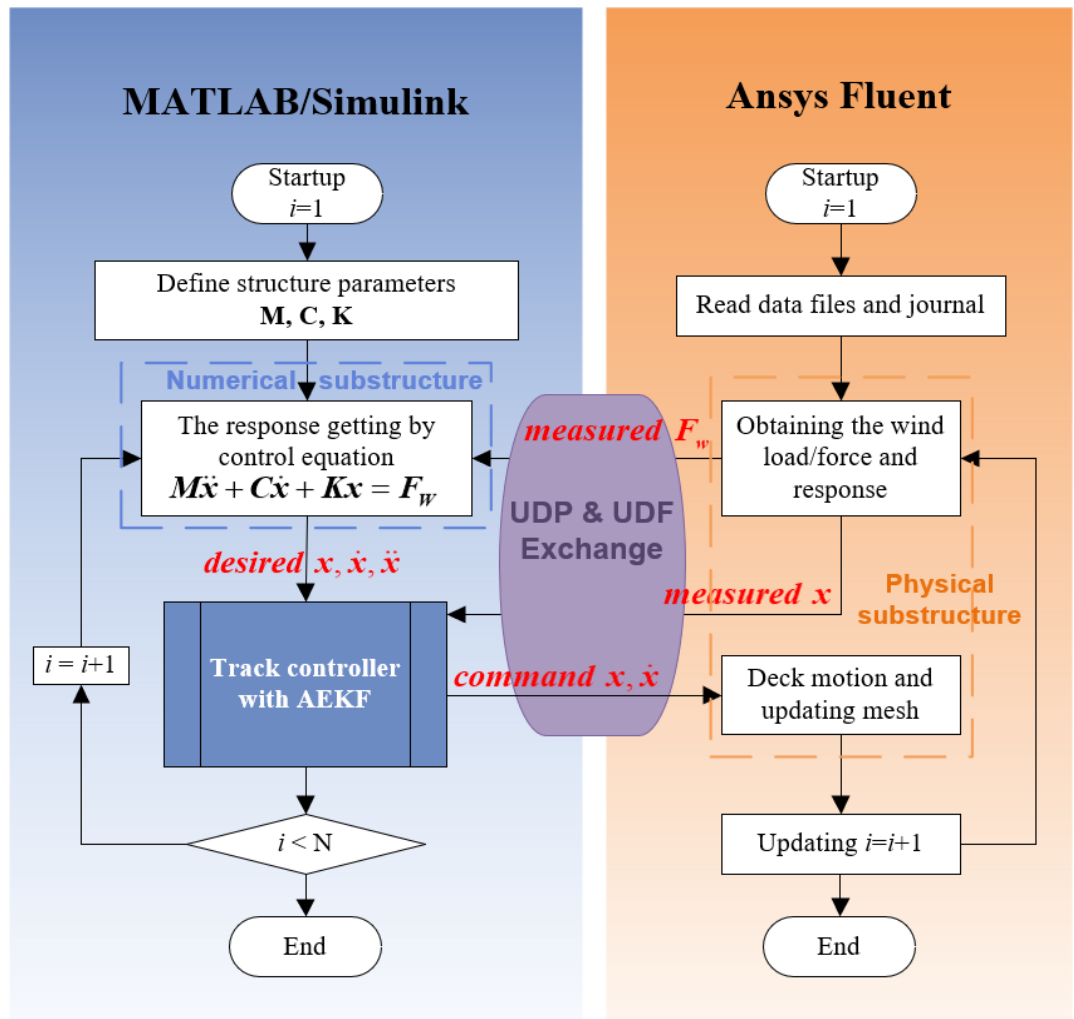

Fig. 10 The flowchart of the MATLAB/Simulink – FLUENT/UDF co-simulation framework

The data exchange between the two platforms is implemented via User Datagram Protocol (UDP) and User-Defined Function (UDF), which is attached in the supplementary material of this paper. MATLAB/Simulink sends data to Fluent UDP, which adds a header containing the source and destination port numbers, forming a UDP datagram. This datagram is then passed to the IP protocol of the network layer, which is responsible for routing and transmitting the data. Upon receipt, the receiver's UDP forwards the datagram to Fluent according to the specified port number (Thuneibat and Al Sharaa, 2023). Notably, UDP does not ensure the order of data arrival or handle data loss; however, its transmission speed is very fast, making it ideal for applications with high real-time requirements (Machado, 2022). Additionally, UDP's excellent multicast capabilities enable fluent parallel computing scheduling. Its flexibility allows developers to control transmission details, such as custom error detection and correction mechanisms to mitigate data loss.

UDF is a tool that allows users to write custom functions to extend the capabilities of databases or data processing systems. In this model, UDF in Fluent receives and sends data to MATLAB through UDP, extracts model signals, sets body motion, and updates the mesh through specified functions. The co-simulation framework leverages the strengths of Fluent for flow field solutions and MATLAB for data processing. MATLAB's extensive toolbox can complement Fluent's C language-based UDF for handling data processing tasks. Therefore, this co-simulation model can efficiently handle problems such as multi-physics field coupling or control algorithm design.

*4.3.2 Numerical validation of 2DOF system*

To further verify the applicability of the RTAHS in section model wind tunnel tests, numerical validation of the heave-torsion coupled system is considered. The governing equations for the two-DOF system are given as

$$m(\ddot{h} + 2\xi_h \omega_h \dot{h} + \omega_h^2 h) = L_{se}(U, \omega, \dot{h}, \dot{\alpha}, \alpha, h) \tag{16a}$$

$$I(\ddot{\alpha} + 2\xi_\alpha \omega_\alpha \dot{\alpha} + \omega_\alpha^2 \alpha) = M_{se}(U, \omega, \dot{h}, \dot{\alpha}, \alpha, h) \tag{16b}$$

where $\ddot{\alpha}$, $\dot{\alpha}$ and $\alpha$ are the torsional acceleration, velocity and displacement, respectively. $I$ is the equivalent mass moment of inertia per unit length of the section model oscillatory system; $L_{se}$ is the aeroelastic lift force and $M_{se}$ is the aeroelastic torsional moment per unit length.

Eq.(16) is then transformed into the space state form as

$$\boldsymbol{x} = \begin{bmatrix} h & \dot{h} & \alpha & \dot{\alpha} \end{bmatrix}^T; \boldsymbol{y} = \begin{bmatrix} h & \alpha \end{bmatrix}^T; \boldsymbol{u} = \begin{bmatrix} L_{se} & M_{se} \end{bmatrix}^T \tag{17}$$

Referring to Eq.(4), the matrices $\boldsymbol{A}$, $\boldsymbol{B}$ and $\boldsymbol{H}_m$ are defined as follows

$$\boldsymbol{A} = \begin{bmatrix} 0 & 1 & & \\ \omega_h^2 & -2\xi_h \omega_h & & \\ & & 0 & 1 \\ & & \omega_\alpha^2 & -2\xi_\alpha \omega_\alpha \end{bmatrix}; \boldsymbol{B} = \begin{bmatrix} 0 & \\ 1/m & \\ & 0 \\ & 1/I \end{bmatrix}; \boldsymbol{H}_m = \begin{bmatrix} 1 & 0 & & \\ & & 1 & 0 \end{bmatrix} \tag{18}$$

This constitutes the calculation logic of the RTAHS numerical substructure. The aerodynamic force required during the calculation process is provided in real-time by Fluent. The numerical substructure based on AEKF is constructed in Simulink, as shown in Fig.11. In Simulink, data is received and sent to Fluent via the *UDP receive* and *UDP send* blocks. Since the UDP protocol commonly supports the unit8 data type, the *Byte Unpack* block and *Byte Pack* block should be inserted for conversion between a uint8 byte vector and multiple vectors of varying data types. Data conversion operations are also required in Fluent. The AEKF is also implemented through functions in Simulink.

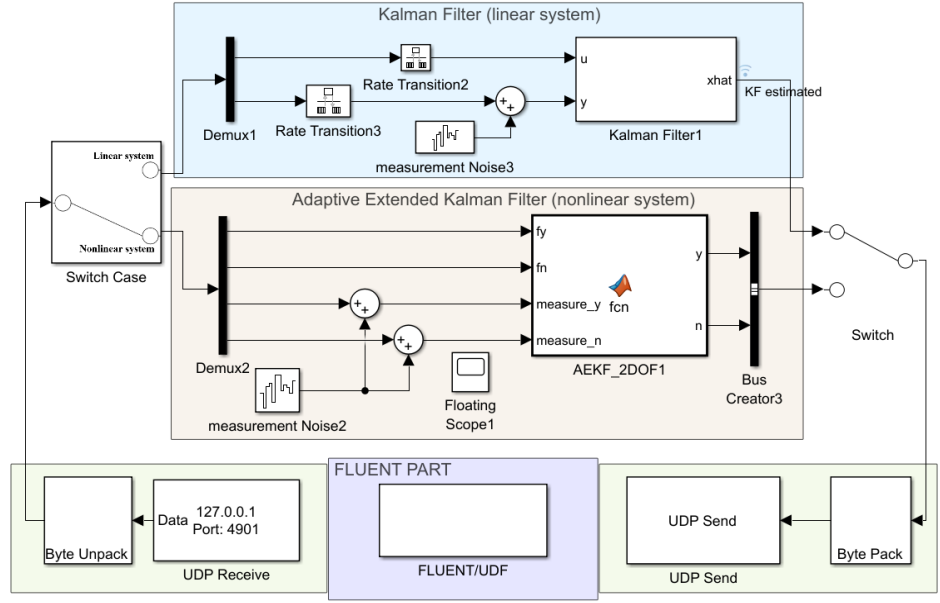
Fig. 11 MATLAB/Simulink–FLUENT/UDF co-simulation framework.

The co-simulation framework and RTAHS method are validated in a two-DOF system of a long-span suspension bridge (Larsen and Jacobsen, 2017). The model parameters are listed in Table 3. RANS simulations were employed to model the wind-structure interaction. And the simulations were carried out using the SST k-ω turbulence model, with an inflow turbulence intensity of 0.5% and a turbulent viscosity ratio of 10%. The velocity-pressure coupling was solved using the SIMPLE algorithm, a second-order implicit scheme was utilized in temporal discretization, while spatial discretization employed the Second-order Gaussian scheme. The calculation time step was 0.001s, and the convergence residual was $10^{-5}$.

The numerical results of the two-DOF system are displayed in Fig. 12, where the calculated time histories under two typical reduced velocities are provided. The results of the proposed AEKF-RTAHS are compared with those by CFD UDF alone, which can be treated as the equivalence of the section model tests in wind tunnel. By examining the curves of displacement, it can be found that the results obtained from the proposed AEKF-RTAHS are in good agreement with those from CFD alone. Therefore, the feasibility and effectiveness of the proposed RTAHS framework as well as the MATLAB/Simulink – FLUENT/UDF co-simulation framework are verified.

Table 3 Model parameters of a long-span bridge (Larsen and Jacobsen, 2017)

| Parameters (unit) | Prototype bridge | Section model |
| --- | --- | --- |
| Geometric scale ratio | | 1:50 |
| Wind speed scale ratio | | 1:6 |
| Length (m) | 1624 | - |
| Width (m) | 31 | 0.62 |
| Height (m) | 4.4 | 0.088 |
| Mass per unit length (kg) | $22.74 \times 10^3$ | 9.096 |
| Moment of inertia per unit length (kg·m) | $2.47 \times 10^6$ | 0.3952 |
| Vertical frequency (Hz) | 0.100 | 0.8333 |
| Torsional frequency (Hz) | 0.278 | 2.3166 |
| Damping ratio $\xi$ | | 0.003 |

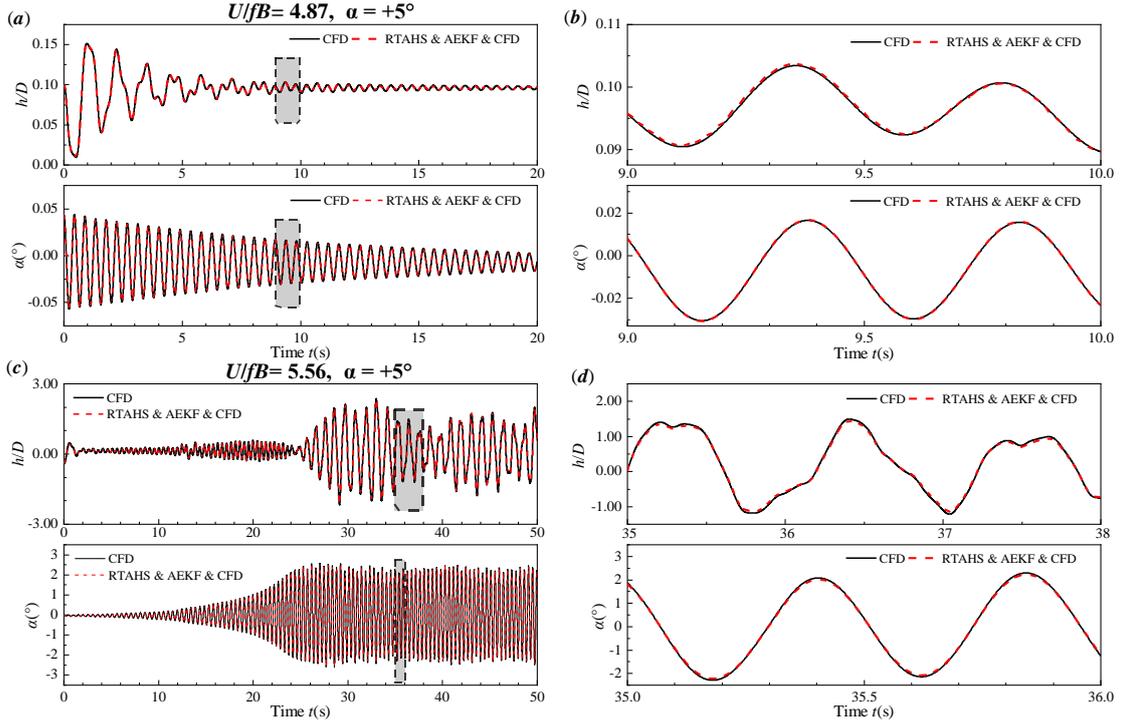

Fig. 12 Numerical validation of a two-DOF system: (a) time histories of vertical and torsional displacement under $U/fB = 4.87$; (b) zoomed-in view of displacement under $U/fB = 4.87$; (c) time histories of vertical and torsional displacement under $U/fB = 5.56$; (d) zoomed-in view of displacement under $U/fB = 5.56$.

## 5. Assessment of time delay

In real-time hybrid simulation, it takes a certain amount of time for the actuator to implement the command sent by the numerical substructure, which will cause the actuator response to have a certain time lag relative to the actuator command, namely the time-delay effect. This time delay equivalently introduces negative damping to the structure (Horiuchi et al., 1999), which can reduce the accuracy of real-time hybrid simulation and even lead to instability in the experiment. Furthermore, the transmission of sensor signals could also incur a time delay. This delay affects the phase relationship between the force and displacement signals, thereby influencing the effective aerodynamic damping. To ensure that the command signal is accurately implemented, inspection of time-delay issue is necessary.

Various time-delay compensation methods have been proposed by scholars, which can be broadly classified into predictive and adaptive approaches (Strano et al., 2022). Horiuchi et al. (1999) employed a polynomial fitting method to approximate the actuator trajectory and extrapolate the command signal, sending several steps of the signal in advance to compensate for the delay. However, this method struggles to ensure stability for structures with short periods. Zhang et al. (2016) incorporated the Kalman filter into the RTHS loop to mitigate high-frequency sensor noise and avoid time delay relative in identifying table trajectory and measurement. Their results demonstrated that the proposed KF-RTHS was stable and reliable, even for large control-structure interactions and low-damping structures. Wang et al. (2024) proposed unscented Kalman filter-based two-stage adaptive compensation (UKF-TAC) for real-time hybrid simulation, which could solve the time-delay problem caused by the servo-hydraulic actuator.

The time delay in the proposed RTAHS of this study consists mainly of two components: mechanical device delay and sensor signal transmission delay. The mechanical delay can be mitigated using rotor position compensation techniques, which are commonly applied in mechanical engineering (Abadía et al., 2021). In the experiment, a noticeable delay between the estimated and actual positions was observed, primarily due to phase delay induced by an RC low-pass filter. To address this issue, phase compensation can be implemented in the software and the cutoff frequency of the low-pass filter can be adjusted in the control hardware to counteract the phase delay. The effectiveness will be further investigated together with the fabricating the proposed mechanical devices in Fig.5.

Sensor signal transmission is highly variable and often subject to delays induced by environmental factors. Compared to laser displacement sensors, piezoelectric accelerometers used in wind tunnel tests are more sensitive to temperature changes, resulting in more significant delay effects. Zhu et al. (2013) observed that force sensors exhibit the largest delay, reaching up to 0.06 s, followed by accelerometers, while laser displacement sensors experience the smallest delay. In this verification section, the delay of the force signal relative to the displacement signal was artificially amplified to 0.1 s to assess the effectiveness of the AEKF algorithm in compensating the time delay.

As illustrated in Fig. 13, with a time delay up to 0.1 seconds, the simulation result closely align with the reference value. This indicates that the RTAHS in this system prioritizes the posterior estimate over the prior one. In this way, significant delays in the force signal can be effectively mitigated, achieving the desired outcome for the experiment. By utilizing the proposed RTAHS with the AEKF algorithm, the time-delay stability of the system can be significantly enhanced.

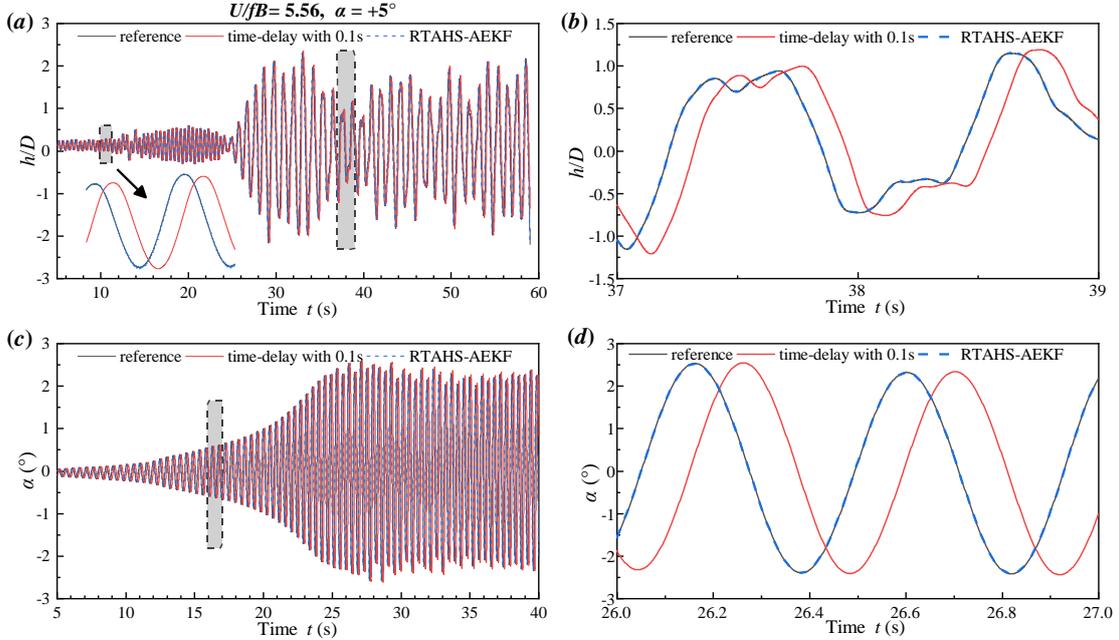

Fig. 13 Numerical testing of time-delay issues: (a) vertical time history response; (b) zoomed-in view of figure a; (c) torsional time history response; (d) zoomed-in view of figure c.

# 6. Conclusions

The conventional experimental technique using spring-suspension system has intrinsic limitations in modelling nonlinear structural behaviors and efficient adjustment of structural parameters. To address these issues, this study develops the idea of real-time hybrid simulation in the section model wind tunnel testing. The fundamental concept involves simulating the elastic supporting system of the flexible structures such as mass, stiffness, and damping properties, via an active control system. In the wind tunnel test, only the aerodynamic shape of the bridge section needs to be accurately replicated.

A novel RTAHS framework is then proposed for the 3DOF heave-transverse-torsion coupled vibration of section model tests. The numerical substructure comprises motion state estimation and a track controller, while the physical substructure consists of a motion control device and a tested section model. The model's movement is governed by the servo motors of the control apparatus, which follow a displacement pattern. The RTAHS is capable of simulating nonlinear effects related to stiffness and damping, as well as three-DOF heave-transverse-torsion coupled vibrations.

The integration of the Adaptive Extended Kalman Filter (AEKF) into the RTAHS loop enables adaptive adjustment and noise filtering, effectively addressing the weak nonlinear issues encountered in wind tunnel tests. Moreover, the AEKF algorithm is capable of mitigating sensor-induced time delay problems. The feasibility and accuracy of the proposed framework were validated through single-DOF and two-DOF systems.

This study also developed a MATLAB/Simulink – FLUENT/UDF co-simulation framework for the rigorous validation of the proposed RTAHS. This framework combines the advantages of MATLAB in data processing and Fluent in flow field solutions. This collaborative simulation approach allows for more efficient design and analysis.

There still are a few limitations in the RTAHS framework which warrant further study. The control device designed in this study still falls short of actual performance, as it lacks physical validation and more robust verification. To address the actuator-induced time delays, an Adaptive Delay Compensation Extended Kalman Filter (AEKF-ADC) algorithm suitable for the RTAHS is under investigation. This can be achieved by implementing a low-pass filter in the feedforward controller to mitigate the impact of delay. Additionally, in the presence of more complex wind-induced responses, the employed algorithms will need refinement to meet the challenges posed by various experimental conditions, including aeroelastic stability and multimodal coupling effects (Gao et al., 2025).

# Acknowledgments

The funding was provided by National Natural Science Foundation of China, 52278478, by the Research Funds for the Interdisciplinary Projects, CHU (No. 300104240923), Fundamental Research Funds for the Central Universities, CHD (No. 300102214914).

# Appendix: Deduction of the Adaptive Extended Kalman Filter algorithm

The governing equation of motion for a single-DOF system can be written as:
$$M\ddot{x}(t) + C\dot{x}(t) + Kx(t) = F(t) \tag{A-1}$$

where $M$, $C$, $K$ are the structural mass, damping and stiffness, respectively; $F(t)$ is the unknown external force.

For a single-DOF nonlinear system, its state space equation and observation equation can be expressed as:
$$x_k = \mathbf{f}(x_{k-1}, u_{k-1}, w_{k-1}) \tag{A-2}$$
$$z_k = \mathbf{h}(x_k, v_k) \tag{A-3}$$

Where subscript $k$ refers to the $k$th time instant with $t = k \cdot \delta t$, and $\delta t$ is the sampling time interval; $x_k$ is the state vector; $\mathbf{f}(\cdot)$ refers to the state transfer function, which obeys the first-order Markov assumption; $u_{k-1}$ is system input vector; $w_{k-1}$ represents the system process noise vector with covariance matrix $Q_{k-1}$; $z_k$ is the observation vector; $\mathbf{h}(\cdot)$ refers to the observation function of the system, subject to the observation independence assumption; $v_k$ represents the measurement noise vector with covariance matrix $R_k$.

The Taylor series expansion method is adopted to linearize the nonlinear system. $x_k$ and $z_k$ expand around the posterior state estimate $\hat{x}_{k-1}$ at the $(k-1)$th time and the prior state estimate $\hat{x}_{k|k-1}$ at the $k$th time, respectively. The high-order terms of the Taylor series expansion are ignored and obtaining the linearized state space equation by partial derivatives process:

$$x_k = \mathbf{A}_k x_{k-1} + \mathbf{W}_k w_{k-1} + (\hat{x}_{k|k-1} - \mathbf{A}_k \hat{x}_{k-1}) \tag{A-4}$$
$$z_k = \mathbf{H}_k x_k + \mathbf{V}_k v_k + (\mathbf{h}(\hat{x}_{k|k-1}, v_k) - \mathbf{H}_k \hat{x}_{k|k-1}) \tag{A-5}$$
$$\mathbf{A}_k = [\partial \mathbf{f}(\hat{x}_{k-1}, u_{k-1}, w_{k-1}) / \partial x] \tag{A-6}$$
$$\mathbf{W}_k = [\partial \mathbf{f}(\hat{x}_{k-1}, u_{k-1}, w_{k-1}) / \partial w] \tag{A-7}$$
$$\mathbf{H}_k = [\partial \mathbf{h}(\hat{x}_{k|k-1}, v_k) / \partial x] \tag{A-8}$$
$$\mathbf{V}_k = [\partial \mathbf{h}(\hat{x}_{k|k-1}, v_k) / \partial v] \tag{A-9}$$

where $(\hat{x}_{k|k-1} - \mathbf{A}_k \hat{x}_{k-1})$ and $(\mathbf{h}(\hat{x}_{k|k-1}, v_k) - \mathbf{H}_k \hat{x}_{k|k-1})$ are two calculable vectors at $k$th time, which can be understood as the deviation. They will become zero in the partial derivatives process since they are constant vectors; $\mathbf{A}_k$ and $\mathbf{W}_k$ are the Jacobian matrices of the partial derivatives of $\mathbf{f}(\cdot)$ with respect to $x$ and $w$, respectively; $\mathbf{H}_k$ and $\mathbf{V}_k$ are the Jacobian matrices of the partial derivatives of $\mathbf{h}(\cdot)$ with respect to $x$ and $v$, respectively. At this time, the process noise and measurement noise follow the normal distribution respectively: $\mathbf{W}_k w_{k-1} \sim N(\hat{q}_{k-1}, \mathbf{W}_k Q_k \mathbf{W}_k^T)$; $\mathbf{V}_k v_k \sim N(\hat{r}_{k-1}, \mathbf{V}_k R_k \mathbf{V}_k^T)$.

Next, the extended Kalman filter continuously and recursively estimates the system state through two key steps: prediction and updation. Eq.(A-10) to Eq.(A-14) represent the five processes of state prediction, error covariance prediction, solving filter gain, state update, and error covariance update, respectively.

$$\hat{x}_{k|k-1} = \mathbf{f}(\hat{x}_{k-1}, u_{k-1}, w_{k-1}) \tag{A-10}$$

$$\mathbf{P}_{k|k-1} = \mathbf{A}_k \mathbf{P}_{k-1} \mathbf{A}_k^{\mathrm{T}} + \mathbf{W}_k \mathbf{Q}_k \mathbf{W}_k^{\mathrm{T}} \tag{A-11}$$

$$\mathbf{K}_k = \mathbf{P}_{k|k-1} \mathbf{H}_k^{\mathrm{T}} \left[ \mathbf{H}_k \mathbf{P}_{k|k-1} \mathbf{H}_k^{\mathrm{T}} + \mathbf{V}_k \mathbf{R}_k \mathbf{V}_k^{\mathrm{T}} \right]^{-1} \tag{A-12}$$

$$\hat{x}_k = \hat{x}_{k|k-1} + \mathbf{K}_k \left[ z_k - \mathbf{h}(\hat{x}_{k|k-1}, v_k) \right] \tag{A-13}$$

$$\mathbf{P}_k = \left[ \mathbf{I} - \mathbf{K}_k \mathbf{H}_k \right] \mathbf{P}_{k|k-1} \tag{A-14}$$

where $\hat{x}_{k|k-1}$ and $\mathbf{P}_{k|k-1}$ are the prior state estimate and prior error covariance estimate at the $k$th time; $\mathbf{K}_k$ is Kalman Gain matrix; $\hat{x}_k$ and $\mathbf{P}_k$ are posterior estimates accordingly.

Compared with the extended Kalman filter, the adaptive extended Kalman filter introduces an adaptive noise estimator, which improves the accuracy of state estimation by adaptively updating the covariance of measurement and process noise. This operation can be named covariance matching and its equations are shown below.

$$\boldsymbol{\varepsilon}_k = z_k - \mathbf{H}_k \hat{x}_{k|k-1} - \hat{r}_{k-1} \tag{A-15}$$

$$\hat{q}_k = (1 - d_k)\hat{q}_{k-1} + d_k(\hat{x}_k - \mathbf{A}_k \hat{x}_{k-1}) \tag{A-16}$$

$$\hat{Q}_k = (1 - d_k)\hat{Q}_{k-1} + d_k(\mathbf{K}_k \boldsymbol{\varepsilon}_k \boldsymbol{\varepsilon}_k^{\mathrm{T}} \mathbf{K}_k^{\mathrm{T}} + \mathbf{P}_k - \mathbf{A}_k \mathbf{P}_{k-1} \mathbf{A}_k^{\mathrm{T}}) \tag{A-17}$$

$$\hat{r}_k = (1 - d_k)\hat{r}_{k-1} + d_k(z_k - \mathbf{H}_k \hat{x}_{k|k-1}) \tag{A-18}$$

$$\hat{R}_k = (1 - d_k)\hat{R}_{k-1} + d_k(\boldsymbol{\varepsilon}_k \boldsymbol{\varepsilon}_k^{\mathrm{T}} - \mathbf{H}_k \mathbf{P}_{k|k-1} \mathbf{H}_k^{\mathrm{T}}) \tag{A-19}$$

where $\boldsymbol{\varepsilon}_k$ is the residual between the measurement value in the physical substructure and the predicted value in the numerical substructure at each step; $d_k = (1-b)/(1-b^k)$; $b$ is a forgetting factor ranging between 0.95 and 0.995, and it is set to 0.96 in this paper.

## CRediT authorship contribution statement

**Wenkai Du**: Writing-Original Draft, Software, Validation, Formal analysis, Conceptualization, Visualization, Methodology. **Guangzhong Gao**: Conceptualization, Methodology, Writing-Editing, Funding acquisition. **Suhan Li**: Validation, Writing-Review. **Bo Fu**: Validation, Writing-Review, Resources. **Jiawu Li**: Supervision, Methodology, Writing-Review. **Ledong Zhu**: Conceptualization, Methodology, Supervision.

## Supplementary Material

The framework code is released in the form of an attachment.

## Nomenclature

| | | | |
|---|---|---|---|
| $a$ | transient amplitude (m) | $\mathbf{P}_{k\|k-1}, \hat{\mathbf{x}}_{k\|k-1}$ | prior error covariance and state estimate |
| $\mathbf{A}$ | state matrix | $\mathbf{P}_k, \hat{\mathbf{x}}_k$ | posterior error covariance and state estimate |
| $b$ | forgetting factor (-) | $\hat{\mathbf{q}}_k$ | process noise mean vector |
| $\mathbf{B}$ | input matrix | $\mathbf{Q}_k$ | process noise covariance matrix |
| $c$ | total damping (-) | $\hat{\mathbf{r}}_k$ | measurement noise mean vector |
| $c_a$ | non-wind-induced additional damping (-) | $\mathbf{R}_k$ | measurement noise covariance matrix |
| $c_s$ | structural damping (-) | $\mathbf{u}$ | input vector relative to wind-induced force |
| $\mathbf{C}_s$ | damping matrix of the spring-suspended system | $U$ | oncoming wind speed (m/s) |
| $\mathbf{C}$ | total damping matrix | $V(t)$ | vertical position at $t$ time (m) |
| $D$ | height of the section model (m) | $\mathbf{v}$ | measurement noise vector |
| $f_m$ | total force between the section model and spring-suspension system (N) | $\mathbf{w}$ | system process noise vector |
| $f_{VI}$ | vortex-induced vertical force (N) | $x, \dot{x}, \ddot{x}$ | response in single-DOF (m, m/s, m/s$^2$) |
| $f_w$ | wind-induced force in flowing air (N) | $\mathbf{x}, \dot{\mathbf{x}}$ | state vector |
| $f_{w0}$ | non-wind-induced additional forces (N) | $\mathbf{y}$ | observation vector of physical model |
| $\mathbf{f}(\cdot)$ | state transfer function | $Y_i, \varepsilon$ | aerodynamic parameter (-) |
| $\mathbf{F}_m$ | force matrix between the section model and spring-suspension system | $\alpha, \dot{\alpha}, \ddot{\alpha}$ | torsional response (rad, rad/s, rad/s$^2$) |
| $\mathbf{F}_w$ | vector of wind-induced force | $\delta H$ | lateral displacement (m) |
| $h, \dot{h}, \ddot{h}$ | vertical response of deck model (m, m/s, m/s$^2$) | $\delta l$ | moving displacement of actuator (m) |
| $\mathbf{h}(\cdot)$ | observation function | $\delta V$ | vertical displacement (m) |
| $H(t)$ | horizontal position at $t$ time (m) | $\delta\alpha$ | rotation angle (rad) |
| $\mathbf{H}$ | observation matrix | $\delta\beta$ | target displacement of joint motor (rad) |
| $I$ | effective inertia moment per unit length (kg·m$^2$/m) | $\delta\theta$ | target displacement of section model (rad) |
| $k$ | spring stiffness (N/m) | $\rho$ | air density (kg/m$^3$) |
| $\mathbf{K}$ | stiffness matrix | $\omega_{h0}$ | circular frequency of vertical bending under small amplitude (rad/s) |
| $\mathbf{K}_k$ | Kalman gain matrix | $\omega_h, \omega_\alpha$ | circular frequency (rad/s) |
| $L_{se}$ | self-induced force by vertical motion (N/m) | $\omega_{vs}$ | circular frequency of vortex shedding (rad/s) |
| $m$ | total effective mass (kg/m) | $\omega(a)$ | amplitude-dependent circular frequency (rad/s) |
| $m_a$ | non-wind-induced additional mass (kg/m) | $\xi_{h0}$ | heave damping ratio under small amplitude (-) |
| $m_m$ | section model mass (kg) | $\xi_h, \xi_\alpha$ | heaving and torsional damping ratio (-) |
| $m_s$ | spring-suspension system mass (kg) | $\xi(a)$ | amplitude-dependent damping ratio (-) |
| $M_{se}$ | self-induced torsional moment (N·m/m) | $\psi$ | phase difference between the vortex shedding and the displacement response (-) |
| $\mathbf{M}$ | total mass matrix | $\varepsilon_k$ | residual between the measurement value and the predicted value (-) |

Abbreviations:

RTAHS: Real-Time Aeroelastic Hybrid Simulation

DOF: Degree of freedom; KF: Kalman Filter

EKF: Extended Kalman Filter

AEKF: Adaptive Extended Kalman Filter

UDP: User Datagram Protocol

CFD: Computational Fluid Dynamics